\documentclass{ifacconf}
\usepackage{graphicx}      % include this line if your document contains figures
\graphicspath{{figures/}}  % Figures path
\usepackage{tikz} %latex graphics
\usetikzlibrary{calc,intersections,arrows,shapes,positioning}
\usepackage[american,nooldvoltagedirection]{circuitikz}
\usepackage[strings]{underscore}% Control the behavior of "_" in text
\usepackage{textcomp}      % Text companion fonts
\usepackage{microtype}     % Subliminal typographical refinements
\usepackage{natbib}        % required for bibliography
\usepackage{siunitx}	   % A comprehensive (SI) units package
\usepackage{mathtools}	   % Extends amsmath package from the AMS
\usepackage{IEEEtrantools} % IEEE transactions tools
\usepackage{amssymb}	   % Additional mathematical fonts and symbols from the AMS
\allowdisplaybreaks		   % Allows page break for equations
\newcommand{\T}{\mathsf{T}}% Matrix transpose
\usepackage{xspace}        % Inserta "eaten" spaces
\let\OldTexttrademark\texttrademark
\renewcommand{\texttrademark}{\OldTexttrademark\xspace}%
\usepackage{acro}          % Acronyms support
\DeclareAcronym{mpc}{short=MPC,long=model predictive control}
\DeclareAcronym{lcmpc}{short=LCMPC,long=limit cycle model predictive control}
\DeclareAcronym{apf}{short=APF,long=active power filter}
\DeclareAcronym{pcc}{short=PCC,long=point of common coupling}
\DeclareAcronym{thd}{short=THD,long=total harmonic distortion}       % List of acronyms
\usepackage{siunitx}       % SI units

\begin{document}
\begin{frontmatter}

\title{An Approach to State Signal Shaping by\\ Limit Cycle Model Predictive Control\thanksref{footnoteinfo}} 
\thanks[footnoteinfo]{This contribution was partly developed within the project NEW~4.0~(North German Energy Transition~4.0) which is funded by the German Federal Ministry for Economic Affairs and Energy~(BMWI).}
\author[First,Second,Third]{Carlos Cateriano Yáñez} 
\author[First]{Gerwald Lichtenberg}
\author[Second]{Georg Pangalos} 
\author[Third]{Javier Sanchis Sáez}
\address[First]{Life Sciences, Hamburg University of Applied Sciences, Germany\\
\mbox{(email:\{carlos.caterianoyanez;gerwald.lichtenberg\}@haw-hamburg.de).}}
\address[Second]{Fraunhofer Institute for Silicon Technology ISIT, Germany\\
(e-mail: georg.pangalos@isit.fraunhofer.de).}
\address[Third]{Instituto U. de Automática e Informática Industrial, Universitat Politècnica de València, Spain \mbox{(email: jsanchis@isa.upv.es).}}

\begin{abstract}                % Abstract of not more than 250 words.
A novel nonlinear model predictive control approach for state signal shaping is proposed. The control strategy introduces a residual shape cost kernel based on the dynamics of circular limit cycles from a supercritical Neimark-Sacker bifurcation normal form. This allows the controller to impose a fundamental harmonic state signal shape with a specific frequency and amplitude. An application example for harmonic compensation in distribution grids integrated with renewable energies is presented. The controller is tasked with the calculation of the reference current for an active power filter used for load compensation. The results achieved are successful, reducing the harmonic distortion to satisfactory levels while ensuring the correct frequency and amplitude.
\end{abstract}


\end{frontmatter}

%===============================================================================

\section{Introduction}\label{sec:int}
Typically,~\ac{mpc} algorithms are used for reference following control problems, due to their handling of constraints and efficient solving methods in the linear case~\citep{Ma02}. However, there are applications where the state dynamics need to preserve a certain shape, rather than following a reference over time, e.g.\@ multiple robots formation coordination in robotics~\citep{EgXi01}. These kinds of problems in the signal processing area are often referred to as signal shaping~\citep{BrKe14}.

Recently, an~\ac{mpc} approach was developed for linear state signal shaping in~\cite{CaPaLi18}. The proposed controller was used to mitigate the harmonic distortion in an electric grid by defining a harmonic linear \emph{shape class} to control the shape of the system states. While the frequency and shape were controlled, the amplitude of the signals was out of reach, due to the linear limitation of the shape class. An extension to this approach as a quadratic program is presented by~\cite{WeCaPa20}, where the amplitude is successfully constraint but not set to the desired value. This paper proposes a~\ac{lcmpc}, a nonlinear~\ac{mpc} approach that aims to directly control the amplitude using a limit cycle kernel as a nonlinear shape class, while also maintaining the correct shape and frequency.

The~\ac{lcmpc} in contrast to a standard tracking~\ac{mpc}, does not require a time varying-reference to be calculated a priori for the optimization. For the~\ac{lcmpc}, the ``reference'' is the limit cycle autonomous system, which is embedded into the cost function, thus dynamically generated within.

This paper is organized as follows. Section~\ref{sec:perorbs} introduces periodic orbits, particularly limit cycles. In section~\ref{sec:lcmpc}, the~\ac{lcmpc} is developed. Section~\ref{sec:appex} gives an application example for harmonic compensation. Finally, section~\ref{sec:concl} draws conclusions.
\section{Periodic Orbits}\label{sec:perorbs}
Consider a non-constant solution~$x(t)$ of a dynamical system. Such a solution is considered to be periodic if there is a constant~$T>0$, such that
\begin{IEEEeqnarray}{rCl}
x\left(t\right)&=&x\left(t+T\right)\quad \forall t\text{,}\label{eq:Tx}
\end{IEEEeqnarray}
where the minimum possible~$T$ is the period. The image~$x\left(t\right)$ of the periodicity interval~\mbox{$[0, T]$} is called a \emph{periodic orbit}~\citep{IvIv07}.

The following subsections will further explore the properties of periodic orbits of special interest. Subsection~\ref{ssec:limc} introduces the concept of limit cycles. This concept is further extended for maps in subsection~\ref{ssec:maplim}. Finally, subsection~\ref{ssec:scns} introduces a special case of the limit cycle which is the base for the proposed control concept.
%%%%%%%%%%%
\subsection{Limit cycles}\label{ssec:limc}
A periodic orbit~$\mathbf{\Gamma}$ on a plane is called a \emph{limit cycle}, if for some point that is not on the periodic orbit, the limit set is exactly~$\mathbf{\Gamma}$, as time goes forward towards~$+\infty$, known as $\omega$-limit set; or backward towards~$-\infty$, known as $\alpha$-limit set~\citep{IvIv07}. This concept is exemplified by the following system.

Given the vector field
\begin{IEEEeqnarray}{rCl}
\IEEEyesnumber\label{eq:vf}\IEEEyessubnumber*
\frac{\text{d}x_1}{\text{d}t}&=&\alpha_c\mu_c x_1-\omega x_2 -\alpha_c x_1 \left(x_1^2+x_2^2\right)\text{,} \label{eq:vfx1}\\
\frac{\text{d}x_2}{\text{d}t}&=&\omega x_1+\alpha_c\mu_c x_2 -\alpha_c x_2 \left(x_1^2+x_2^2\right)\text{,} \label{eq:vfx2}
\end{IEEEeqnarray}
where~$\left\lbrace\alpha_c,\mu_c,\omega\right\rbrace\in\mathbb{R}_{>0}$ are parameters of the system in~\eqref{eq:vf}. There is a periodic orbit of circular shape that arises from the system~\citep{GuHo83}. This behavior becomes evident when transforming~\eqref{eq:vf} to polar coordinates
\begin{IEEEeqnarray}{rCl}
\IEEEyesnumber\label{eq:hopf}\IEEEyessubnumber*
\frac{\text{d}r}{\text{d}t} & = & \alpha_c r\left(\mu_c-r^2\right)\text{,} \label{eq:hopfr}\\
\frac{\text{d}\theta}{\text{d}t}& = & \omega\text{,}  \label{eq:hopftheta}
\end{IEEEeqnarray}
where the role of the parameters can be easily identified, being~$\alpha_c$ an intensifier parameter,~$\mu_c$ the square radius of the circular orbit, and~$\omega$ the angular frequency in~\si[per-mode=fraction]{\radian\per\second}.

\begin{figure}[ht!]
\centering
\includegraphics[scale=1]{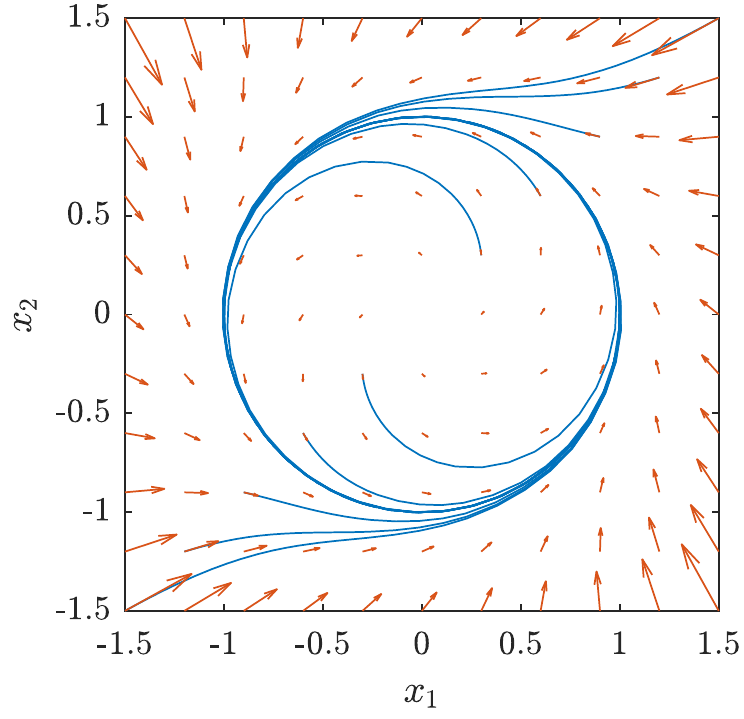}
\caption{Super critical Hopf bifurcation normal form phase portrait with~\mbox{$\alpha_c=\mu_c=\omega=1$}}
\label{fig:phase_hopf}
\end{figure}

An example with all the system parameters set to 1 is shown in the phase portrait in Fig.~\ref{fig:phase_hopf}. From the vector field, an unstable equilibrium at the origin and a stable circular limit cycle of radius 1 arise. The limit cycle is stable because all the system parameters are elements of~$\mathbb{R}_{>0}$, which leads to a supercritical \emph{Hopf bifurcation} normal form~\citep{GuHo83}.

The appearance of the Hopf bifurcation, also known as the \mbox{Poincaré-Andronov-Hopf} bifurcation, occurs when the stability of an equilibrium point in a continuous-time dynamical system changes via a pair of purely imaginary eigenvalues. Hence a limit cycle arises. The bifurcation can be supercritical or subcritical, leading to a stable or unstable limit cycle respectively~\citep{IvIv07}. The normal form used is away from the bifurcation point~\mbox{$\mu_c=0$}. 

This region of attraction of the limit cycle is of special interest in the context of harmonic shape control. By formulating a controller that imposes these dynamics into a second-order plant, it could drive such plant states into a circular limit cycle of choice. This would ensure a fundamental harmonic shape for the plant states of the desired amplitude~$\lvert\sqrt{\mu_c}\rvert$ and angular frequency~$\omega$, regardless of their initial conditions.
%%%%%%%%%%%
\subsection{Limit cycles for maps}\label{ssec:maplim}
In section~\ref{ssec:limc}, the fundamentals of limit cycles were introduced in continuous time. However, to explore its full potential for~\emph{digital} control, an analysis in the discrete-time domain is required.

The normal forms of the \emph{Neimark-Sacker bifurcation}, which is the discrete-time case of the Hopf bifurcation for maps~\citep{DiZh09}, are a suitable starting point for this analysis. 

For discrete time~$t=k\tau$, with sampling time~$\tau$, the normal form in polar coordinates is given as
\begin{IEEEeqnarray}{rCl}
\IEEEyesnumber\label{eq:ns}\IEEEyessubnumber*
r_{k+1}&=& r_k + \mu r_k + \alpha r_k^3\text{,} \label{eq:nsr}\\
\theta_{k+1} & = & \theta_k +\phi\text{,}  \label{eq:nstheta}
\end{IEEEeqnarray}
which consists of a third order simplification of the radius~$r$ map, with parameters~$\mu$ and~$\alpha$, and a linear simplification for the angle~$\theta$ map, with parameter~\mbox{$\phi=\omega\tau$}~\citep{DiZh09}. This simplification was done to match the order of the continuous-time system depicted in~\eqref{eq:hopf}.

Let~\eqref{eq:ns} be transformed into Cartesian coordinates as
\begin{IEEEeqnarray}{rCl}
\mathbf{x}_{k+1} & = & \left( 1+ \mu +\alpha\mathbf{x}_k^\T\mathbf{x}_k\right)\mathbf{R}_{\phi}\mathbf{x}_k\text{,} \label{eq:nscartss}
\end{IEEEeqnarray}
where
\begin{IEEEeqnarray}{rCl}
\IEEEnonumber\IEEEyessubnumber*
\mathbf{x} & = & \begin{bmatrix}x_1&x_2\end{bmatrix}^\T\text{,}\label{eq:nscartssx}\\
\mathbf{R}_{\phi} & = & \begin{bmatrix}\cos\left(\phi\right)&-\sin\left(\phi\right)\\ 
\sin\left(\phi\right)&\cos\left(\phi\right)\end{bmatrix}\text{.}\label{eq:nscartssr}\IEEEnonumber\IEEEyessubnumber*
\end{IEEEeqnarray}
%%%%%%%%%%%%
\subsection{Supercritical Neimark-Sacker bifurcation normal form}\label{ssec:scns}
\begin{figure}[ht!]
\centering
\includegraphics[scale=1]{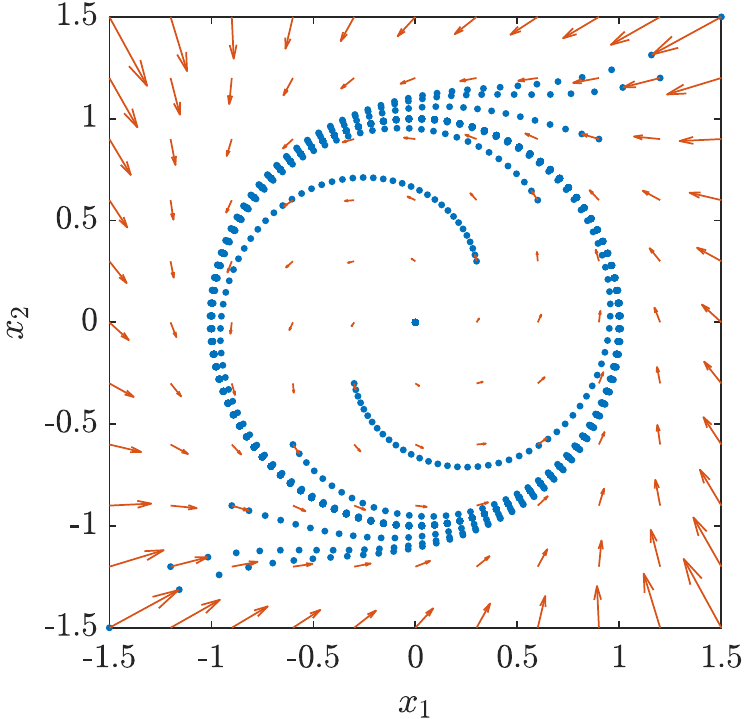}
\caption{Supercritical  Neimark-Sacker bifurcation normal form phase portrait with~\mbox{$\mu=0.05$}, \mbox{$\alpha=-0.05$}, \mbox{$\omega=2\pi50$}, and~\mbox{$\tau=\SI{0.2}{\milli\second}$}}
\label{fig:scns_phase}
\end{figure}

For~$\mu>0$ and~$\alpha<0$ the discrete-time system in~\eqref{eq:nscartss} has an unstable fixed point at the origin and a stable unique circular limit cycle around it~\citep{Ku98}, see Fig.~\ref{fig:scns_phase}. From~\eqref{eq:nsr}, the radius of the circular limit cycle is given by
\begin{IEEEeqnarray}{rCl}
\rho & = & \sqrt{-\frac{\mu}{\alpha}}\text{.} \label{eq:rho}
\end{IEEEeqnarray}

Depending on the radius of a trajectory starting point, different dynamics can be observed. Let
\begin{IEEEeqnarray}{rCl}
\IEEEyesnumber\label{eq:rholim}\IEEEyessubnumber*
\rho_0 & = & \sqrt{-\frac{(1+\mu)}{\alpha}}\text{,} \label{eq:rho0}\\
\rho_\infty & = & \sqrt{-\frac{(2+\mu)}{\alpha}}\text{.} \label{eq:rhoinf}
\end{IEEEeqnarray}

Besides~$r=0$, trajectories starting at~$r=\rho_0$, will lead directly to the unstable fixed point at the origin and remain there unless disturbed. This can be seen by setting~$r=\rho_0$ in~\eqref{eq:nsr}. 

In the case of trajectories starting at~$r>\rho_\infty$, they will tend towards infinity. Once again, this can be observed by setting~$r=\rho_\infty+\epsilon$ in~\eqref{eq:nsr}, where~$\epsilon>0$. In each iteration of the map, the magnitude of~$r$ is increasing to~$\infty^+$, thus~$r$ is divergent.

Therefore, trajectories starting in $0\leq r\leq\rho_\infty$, except for~$r=\rho_0$, will end in the final set given by the circular limit cycle of radius~$\rho$.
\section{Limit Cycle Model Predictive Control}\label{sec:lcmpc}
\begin{figure}[ht!]
\centering
\tikzstyle{block} = [draw, thick, fill=white, rectangle, minimum height=2em, minimum width=3em]
\tikzstyle{sum} = [draw, thick, fill=white, circle,inner sep=1pt,minimum size=0pt, node distance=1.5cm]
\tikzstyle{input} = [coordinate]
\tikzstyle{output} = [coordinate]
\tikzstyle{pinstyle} = [pin edge={to-,thin,black}]
\tikzstyle{connect} = [->,thick]
\tikzstyle{line} = [thick]
\tikzstyle{lineb} = [very thick]
\tikzstyle{connectb} = [->,very thick]
\tikzstyle{lineb} = [very thick]
\tikzstyle{blockb} = [draw, fill=white, rectangle, minimum height=3em, minimum width=4em, text width=4.5em, 
    text centered,very thick]
\tikzstyle{sumb} = [draw,very thick, fill=white, circle,inner sep=1pt,minimum size=0pt, node distance=1.5cm]
\begin{tikzpicture}[auto, node distance=1.75cm,>=latex',scale =1, transform shape]
\node(mpc)[blockb] at (0,0){MPC};
\node [sumb, right =1cm of mpc] (sum1) {};
\node [blockb, right=0.5cm of sum1] (plant) {Plant};
\node [sumb, right =0.5cm of plant] (sum2) {};
\draw [connectb](mpc)--node(u){$\mathbf{u}$}(sum1);
\draw [connectb](sum1)--(plant);
\draw [connectb](plant)--(sum2);
\node [coordinate, below of=mpc](ref){};
\node[above of=sum1](v){$\mathbf{v}$};
\node[above of=mpc](vh){$\hat{\mathbf{v}}$};
\node[above of=sum2](w){$\mathbf{w}$};
\draw [connectb](v)--(sum1);
\draw [connectb](w)--(sum2);
\draw[connectb](sum2.south)--+(0pt,0pt)node[above,xshift=.25cm,yshift=-.5cm]{$\mathbf{y}$}|-(ref)-|( $(mpc.west)+(-0.5cm,0pt)$)--( $(mpc.west)$);
\draw[connectb](vh)--(mpc.north);
\end{tikzpicture}
\caption{\Acl{mpc} closed loop}
\label{fig:mpcloop}
\end{figure}
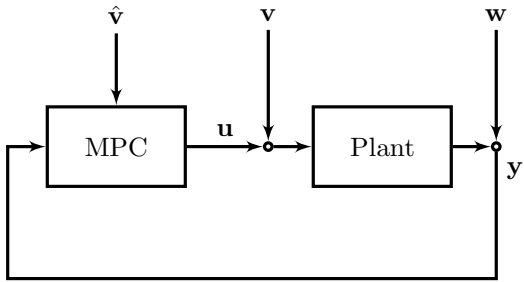

The basis of an~\ac{mpc} is the solution of an optimization problem such as
\begin{IEEEeqnarray}{C}
\min_{\mathbf{Z}} {J\left(\mathbf{Z}\right)}\text{,}\label{eq:minJ}
\end{IEEEeqnarray}
where~$J$ is the cost function and~$\mathbf{Z}$ the set of decision variables~\citep{Ma02}. The cost function is calculated in discrete time for a finite prediction horizon~$H_p\in\mathbb{N}$. In this condensed formulation, the cost function~$J$ also includes model parameters of the target system. The proposed controller targets linear systems, which are governed by the discrete-time state space equations given as
\begin{IEEEeqnarray}{rCl}
\IEEEyesnumber\label{eq:dss}\IEEEyessubnumber*
\mathbf{x}_{k+1}& = & \mathbf{A}\mathbf{x}_{k}+\mathbf{B}\mathbf{u}_{k}+\mathbf{F}\mathbf{v}_{k}\text{,}\label{eq:dssx}\\ 
\mathbf{y}_{k}& = & \mathbf{C}\mathbf{x}_{k}+\mathbf{w}_{k}\text{,} \label{eq:dssy}
\end{IEEEeqnarray}
with~state vector~\mbox{$\mathbf{x}\in\mathbb{R}^n$}, control input~\mbox{$\mathbf{u}\in\mathbb{R}^m$}, measured input disturbance~\mbox{$\mathbf{v}\in\mathbb{R}^d$}, output~\mbox{$\mathbf{y}\in\mathbb{R}^r$}, output disturbance~\mbox{$\mathbf{w}\in\mathbb{R}^r$}, system matrix~\mbox{$\mathbf{A}\in\mathbb{R}^{n\times n}$}, input matrix~\mbox{$\mathbf{B}\in\mathbb{R}^{n\times m}$}, input disturbance matrix~\mbox{$\mathbf{F}\in\mathbb{R}^{n\times d}$}, and output matrix~\mbox{$\mathbf{C}\in\mathbb{R}^{r\times n}$}. Fig.~\ref{fig:mpcloop} shows a general control closed loop for the~\ac{mpc}, where~$\hat{\mathbf{v}}$ is the measured input disturbance prediction, which could differ from~$\mathbf{v}$.

For the~\ac{lcmpc}, the key feature is the integration of the limit cycle dynamics from section~\ref{sec:perorbs}, directly into the core of its cost function formulation, further detail is provided in subsection~\ref{ssec:costkern}. Subsection~\ref{ssec:linpred} addresses the prediction layer with the linear systems prediction equations for states. Finally, subsection~\ref{ssec:inpar} gives an input parameterization scheme, narrowing the solution space.
%%%%%%%%%%%
\subsection{Cost function kernel formulation}\label{ssec:costkern}
In order to integrate the limit cycle dynamics into the core of the cost function, a limit cycle residual cost is defined as a kernel. Taking the state space map in~\eqref{eq:nscartss} as starting point, the kernel in discrete time is formulated as
\begin{IEEEeqnarray}{rCl}
\mathbf{0}_{2\times 1} & = & \mathbf{x}_{k+1}-\left(1+\mu\right)\mathbf{R}_{\phi}\mathbf{x}_k-\alpha\mathbf{R}_{\phi}\mathbf{x}_k \mathbf{x}^\T_k \mathbf{x}_k\text{,} \label{eq:kern}
\end{IEEEeqnarray}
where~$\mathbf{0}_{2\times 1}$ is a vector of zeros in~$\mathbb{Z}^{2\times 1}$.This kernel expression is then squared and accumulated for the whole prediction horizon~$H_p$, leading to the cost function
\begin{IEEEeqnarray}{rCl}
J\left(\mathbf{X}\right)& = & \mathbf{X}^\T\mathbf{Q}_2\mathbf{X}+2\alpha\mathbf{X}^\T\left(\mathbf{L}\circ\left(\mathbf{X}\mathbf{X}^\T\mathbf{Q}_4\right)\right)\mathbf{X}\label{eq:Jvec}\\\nonumber
&&+\alpha^2\mathbf{X}^\T\left(\mathbf{L}\circ\left(\mathbf{X}\mathbf{X}^\T\left(\mathbf{L}\circ\left(\mathbf{X}\mathbf{X}^\T\right)\right)\right)\mathbf{X}\right)\text{,}
\end{IEEEeqnarray}
where
\footnotesize
\begin{IEEEeqnarray}{rCl}
\IEEEnonumber\IEEEyessubnumber*
\mathbf{X}& = & \begin{bmatrix}\mathbf{x}_{k+1}&\mathbf{x}_{k+2}&\cdots&\mathbf{x}_{k+H_p}\end{bmatrix}^\T\text{,}\label{eq:XHp}\\
\mathbf{Q}_2& = & \left[\begin{array}{cccc}
\left(1+\mu\right)^2\mathbf{I}_{2\times 2}&-\left(1+\mu\right)\mathbf{R}^\T_{\phi} & \mathbf{0}_{2\times 2} &\cdots\\
-\left(1+\mu\right)\mathbf{R}_{\phi}&\left(1+\left(1+\mu\right)^2\right)\mathbf{I}_{2\times 2}&\ddots &\ddots\\
\mathbf{0}_{2\times 2}&\ddots &\ddots &\ddots\\
\vdots&\ddots &\ddots &\ddots\\
\vdots&\ddots &\ddots &\ddots\\
\vdots&\ddots &\ddots &\ddots\\
\mathbf{0}_{2\times 2}&\cdots &\cdots &\cdots
\end{array}\right.\nonumber\\
&&\left.\begin{array}{cccc}
\cdots  &\cdots &\cdots & \mathbf{0}_{2\times 2}\\
\ddots & \ddots  &\ddots & \vdots\\
\ddots &\ddots &\ddots & \vdots\\
\ddots &\ddots &\ddots & \vdots\\
\ddots &\ddots &\ddots & \mathbf{0}_{2\times 2}\\
\ddots &\ddots &\left(1+\left(1+\mu\right)^2\right)\mathbf{I}_{2\times 2}& -\left(1+\mu\right)\mathbf{R}^\T_{\phi}\\
\cdots &\mathbf{0}_{2\times 2} &-\left(1+\mu\right)\mathbf{R}_{\phi} & \mathbf{I}_{2\times 2}
\end{array}\right]\text{,} \label{eq:Q2}\\
\mathbf{L}& = & \begin{bmatrix}
\mathbf{1}_{2\times 2}&\mathbf{0}_{2\times 2}& \cdots & \cdots & \mathbf{0}_{2\times 2}\\
\mathbf{0}_{2\times 2}&\ddots & \ddots & \ddots  & \vdots\\
\vdots&\ddots & \ddots & \ddots  & \vdots\\
\vdots & \ddots &\ddots &\mathbf{1}_{2\times 2} &\mathbf{0}_{2\times 2}\\
\mathbf{0}_{2\times 2}&\cdots &\cdots &\mathbf{0}_{2\times 2}&\mathbf{0}_{2\times 2}
\end{bmatrix}\text{,} \label{eq:L}\\
\mathbf{Q}_4& = & \begin{bmatrix}
\left(1+\mu\right)\mathbf{I}_{2\times 2}&\mathbf{0}_{2\times 2}& \cdots & \cdots & \cdots & \mathbf{0}_{2\times 2}\\
-\mathbf{R}_{\phi}&\ddots & \ddots & \ddots  & \ddots & \vdots\\
\mathbf{0}_{2\times 2}&\ddots & \ddots & \ddots & \ddots  & \vdots\\
\vdots&\ddots & \ddots & \ddots & \ddots  & \vdots\\
\vdots & \ddots &\ddots &\ddots &\left(1+\mu\right)\mathbf{I}_{2\times 2}&\mathbf{0}_{2\times 2}\\
\mathbf{0}_{2\times 2}&\cdots &\cdots &\mathbf{0}_{2\times 2}&-\mathbf{R}_{\phi}&\mathbf{0}_{2\times 2}
\end{bmatrix}\!\!\text{,} \label{eq:Q4}
\end{IEEEeqnarray}
\normalsize
with dimensions, \mbox{$\mathbf{X}\!\in\!\mathbb{R}^{2H_p}\!$}, \mbox{$\mathbf{Q}_2\!\in\!\mathbb{R}^{2H_p\times 2H_p}\!$}, \mbox{$\mathbf{L}\!\in\!\mathbb{R}^{2H_p\times 2H_p}\!$}, and~\mbox{$\mathbf{Q}_4\!\in\!\mathbb{R}^{2H_p\times 2H_p}\!$}; where~$\circ$ is the Hadamard product and~\mbox{$\left\lbrace\mathbf{0}_{j\times j},\mathbf{1}_{j\times j},\mathbf{I}_{j\times j}\right\rbrace\!\in\!\mathbb{Z}^{j\times j}_{\geq0}$} are matrices of zeros, ones, and identity respectively.

This nonlinear cost function defines the future state trajectories vector~$\mathbf{X}$ as the set of decision variables of the underlying optimization problem. Finding the optimal decision variable set will ensure the minimum residual limit cycle cost, meaning that the controller should steer the plant such that its future state trajectories~$\mathbf{X}$ follow exactly this optimal set. Therefore, the prediction of~$\mathbf{X}$ is critical to find the correct input sequence to steer the plant, this is further detailed in subsection~\ref{ssec:linpred}.
%%%%%%%%%%%
\subsection{Linear system predictions}\label{ssec:linpred}
From this subsection and on, predictions of variables will be identified by a hat symbol~``$\;\;\hat{\vphantom{.}}\;\;$''. As stated at the end of subsection~\ref{ssec:costkern}, the cost function only requires predictions of the future states~$\hat{\mathbf{x}}$. Therefore, a prediction of the future state trajectories~$\hat{\mathbf{x}}$ up to a prediction horizon~$H_p$, can be iteratively calculated with~\eqref{eq:dssx}. Given the initial states~$\mathbf{x}_k$, considering a prediction for the future measured input disturbances~$\hat{\mathbf{v}}$, and a future input sequence~$\hat{\mathbf{u}}$~\citep{Ma02}; the linear system prediction equations in vector form are
\begin{IEEEeqnarray}{rCl}
\mathbf{X}\left(\mathbf{U}\right)& = & \mathbf{\Psi}\mathbf{x}_k+\mathbf{\Theta}\mathbf{U}+\mathbf{\Gamma}\mathbf{V}\text{,} \label{eq:liftsys}
\end{IEEEeqnarray}
where
\footnotesize
\begin{IEEEeqnarray}{rCl}
\IEEEnonumber\IEEEyessubnumber*
\mathbf{X}& = & \begin{bmatrix}
\hat{\mathbf{x}}_{k+1}&\cdots&\hat{\mathbf{x}}_{k+H_p}\end{bmatrix}^\T
\in\mathbb{R}^{nH_p}\text{,}\label{eq:xpred}\\
\mathbf{\Psi}& = & \begin{bmatrix}\mathbf{A}&\mathbf{A}^2&\cdots&\mathbf{A}^{H_p}\end{bmatrix}^\T\in\mathbb{R}^{nH_p\times n}\text{,}\label{eq:Psi}\\
\mathbf{\Theta}& = & \begin{bmatrix}
\mathbf{B}&\mathbf{0}_{r\times m}&\cdots &\mathbf{0}_{r\times m}\\
\mathbf{A}\mathbf{B}&\mathbf{B}&\cdots &\mathbf{0}_{r\times m}\\
\vdots &\vdots &\ddots &\vdots\\
\mathbf{A}^{H_p-1}\mathbf{B}&\mathbf{A}^{H_p-2}\mathbf{B}&\cdots &\mathbf{B}
\end{bmatrix}\!\!\in\mathbb{R}^{nH_p\times mH_p}\!,\label{eq:Theta}\\
\mathbf{U}& = & \begin{bmatrix}\hat{\mathbf{u}}_{k}&\cdots&\hat{\mathbf{u}}_{k+H_p-1}\end{bmatrix}^\T\in\mathbb{R}^{mH_p}\text{,}\label{eq:upred}\\
\mathbf{\Gamma}& = & \begin{bmatrix}
\mathbf{F}&\mathbf{0}_{r\times d}&\cdots &\mathbf{0}_{r\times d}\\
\mathbf{A}\mathbf{F}&\mathbf{F}&\cdots &\mathbf{0}_{r\times d}\\
\vdots &\vdots &\ddots &\vdots\\
\mathbf{A}^{H_p-1}\mathbf{F}&\mathbf{A}^{H_p-2}\mathbf{F}&\cdots &\mathbf{F}
\end{bmatrix}\!\!\in\mathbb{R}^{nH_p\times dH_p}\!\text{,}\label{eq:Gamma}\\
\mathbf{V}& = & \begin{bmatrix}\hat{\mathbf{v}}_{k}&\cdots&\hat{\mathbf{v}}_{k+H_p-1}\end{bmatrix}^\T\in\mathbb{R}^{dH_p}\text{.}\label{eq:vpred}
\end{IEEEeqnarray}
\normalsize

Using this set of equations, the decision variable~$\mathbf{X}$ can be expressed in terms of the future input sequence vector~$\mathbf{U}$, which in turn becomes the new decision variable of the optimization problem. This allows the controller input action~$\mathbf{u}$ to be directly aligned with the optimization decision variable.
%%%%%%%%%%%
\subsection{Input Fourier parameterization}\label{ssec:inpar}
Assuming a periodic future input sequence~$\hat{\mathbf{u}}$, typical for harmonic compensation applications, it can be approximated for a fixed bandwidth up to the~$h^{\text{th}}$ harmonic as
\begin{IEEEeqnarray}{rCl}
\hat{\mathbf{u}}(k\tau)&\approx &\sum_{n=1}^{h}\mathbf{f}_n\sin\left(n\omega k\tau\right)+\mathbf{g}_n\cos\left(n\omega k\tau\right)\text{,} \label{eq:ukpar}
\end{IEEEeqnarray}
where~$\mathbf{f}_n$ and~$\mathbf{g}_n$ are Fourier coefficients for the~$n^{\text{th}}$ harmonic of the fundamental angular frequency~$\omega$.

This approximation can be extended up to the prediction horizon~$H_p$ in the following vector form
\begin{IEEEeqnarray}{rCl}
\mathbf{U}\left(\mathbf{P}\right)& = & \left(\mathbf{M}\otimes\mathbf{I}_{m\times m}\right)\mathbf{P}\text{,} \label{eq:spar}
\end{IEEEeqnarray}
where
\footnotesize
\begin{IEEEeqnarray}{rCl}
\IEEEnonumber\IEEEyessubnumber*
\mathbf{M}&=&\left[\begin{array}{@{}ccc@{}}
0&\cdots &0\\
\sin\left(\omega \tau\right)&\cdots &\sin\left(h\omega \tau\right)\\
\vdots &\cdots & \vdots\\
\sin\left(\omega \tau\left(H_p\!-\!1\right)\right)&\cdots &\sin\left(h\omega \tau\left(H_p\!-\!1\right)\right)
\end{array}\right\rvert\cdots\nonumber\\
&&\left\lvert\begin{array}{@{}ccc@{}}
1&\cdots &1\\
\cos\left(\omega \tau\right)&\cdots &\cos\left(h\omega \tau\right)\\
\vdots &\cdots & \vdots\\
\cos\left(\omega \tau\left(H_p\!-\!1\right)\right)&\cdots &\cos\left(h\omega \tau\left(H_p\!-1\!\right)\right)
\end{array}\right]\!\!\!\in\!\mathbb{R}^{H_p\times 2h}\!\text{,}\label{eq:fpar}\\
\mathbf{P}& = & \left[ \begin{array}{@{}ccccc@{}}\mathbf{f}_1&\cdots&\mathbf{f}_h\, \vert \, \mathbf{g}_1&\cdots&\mathbf{g}_h\end{array}\right]^\T\in\mathbb{R}^{2mh}\text{,}\label{eq:par}
\end{IEEEeqnarray}
\normalsize
~and $\otimes$ is the Kronecker product.

By this approximation, the future input sequence vector~$\mathbf{U}$ in~\eqref{eq:liftsys} can be parameterized, leading to
\begin{IEEEeqnarray}{rCl}
\mathbf{X}\left(\mathbf{P}\right)& = & \mathbf{\Psi}\mathbf{x}_k+\mathbf{\Theta}\left(\mathbf{M}\otimes\mathbf{I}_{m\times m}\right)\mathbf{P}+\mathbf{\Gamma}\mathbf{V}\text{.}\label{eq:liftsyspar}
\end{IEEEeqnarray}

Under this parameterization, the decision space of the optimization variable~$\mathbf{U}$ is reduced to harmonic combinations up to the~$h^{\text{th}}$ harmonic order that lead to the Fourier coefficients in~$\mathbf{P}$. Therefore,~$\mathbf{P}$ becomes the new decision variable set, thus reformulating the~\ac{lcmpc} nonlinear optimization problem into
\begin{IEEEeqnarray}{C}
\min_{\mathbf{P}} {J\left(\mathbf{P}\right)}\text{.}\label{eq:minJP}
\end{IEEEeqnarray}
\section{Application Example}\label{sec:appex}
In modern electrical power systems, the increasing share of renewable energy sources leads to voltage and frequency fluctuations that greatly affect the power quality of the distribution grid. Moreover, the abundance of switching power electronics in their generation process also introduces harmonic distortion to the distribution grid~\citep{Li17}.

The focus on this application example is the compensation of the harmonic distortion introduced to the distribution grid, as it is suitable for the harmonic dynamics imposed by the proposed controller.

A typical solution to address harmonic distortion is the implementation of an~\ac{apf}. For successful compensation, the selection of a suitable control scheme that steers the~\ac{apf} is key~\citep{RaMiGh08}. This is where the~\ac{lcmpc} comes into action.

This section is structured as follows. Subsection~\ref{ssec:circmod} introduces the electric grid circuit model as the base for the simulations. In subsection~\ref{ssec:nform}, the base model is transformed into normal form. Subsection~\ref{ssec:simset} describes all the simulation parameters and considerations. Finally, section~\ref{ssec:res} presents the results of the simulations.
%%%%%%%%%%%
\subsection{Electric grid circuit Model}\label{ssec:circmod}
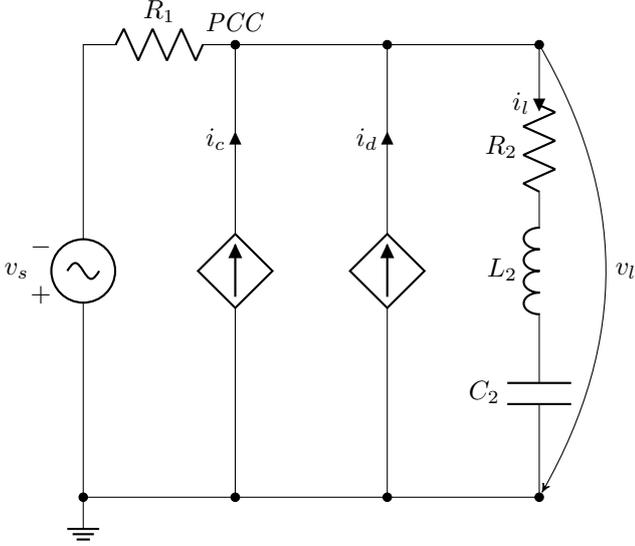
\begin{figure}[ht!]
\centering
\begin{circuitikz}
\draw (0,0)to[sV=$v_{s}$](0,6)to[R,l=$R_{1}$](2,6)node[circ,label={above:\textit{PCC}}]{}--(4,6)node[circ]{}--(6,6)node[circ]{};
\draw(0,0)node[circ]{}--(0,0)node[ground]{};
\draw(0,0)--(2,0)node[circ]{}to[cI=$i_{c}$](2,6);
\draw(2,0)--(4,0)node[circ]{}to[cI=$i_{d}$](4,6);
\draw(4,0)--(6,0)node[circ]{}--(6,.75)to[C=$C_{2}$](6,2)to[L=$L_{2}$](6,4)to[R=$R_{2}$,i^<=$i_l$](6,5.25)--(6,6);
\draw[->,>=stealth',shorten >=2pt](6,6) to [bend left=30]node[midway,right] {$v_{l}$}(6,0);
\end{circuitikz}
\caption{Electric grid circuit model}
\label{fig:gridcirc}
\end{figure}

To show the capabilities of the proposed controller, a minimal electric grid circuit model is used, as shown in Fig.~\ref{fig:gridcirc}. The ideal supply voltage~$v_s$ represents the external distribution grid, the resistance~\mbox{$R_1=\SI{100}{\ohm}$} corresponds to the transmission line between the grid and the~\ac{pcc}. The~\ac{apf} is represented by the ideal controlled current source that feeds in the compensation current~$i_c$. The harmonic distortion is injected into the~\ac{pcc} by the ideal controlled current source that feeds in the disturbance current~$i_d$. Finally, the load current~$i_l$ flows through the series connection of the resistor~\mbox{$R_2=\SI{10}{\ohm}$}, the inductor~\mbox{$L_2=\SI{100}{\milli\henry}$}, and the capacitor~\mbox{$C_2=\SI{10}{\milli\farad}$}, that comprise the load voltage~$v_l$. 

The aforementioned electric grid circuit model is taken into continuous-time state space form as
\begin{IEEEeqnarray}{rCl}
\IEEEyesnumber\label{eq:css}\IEEEyessubnumber*
\mathbf{\dot{x}}\left(t\right)& = & \mathbf{A}_c\mathbf{x}\left(t\right)+\mathbf{b}_c u\left(t\right)+\mathbf{F}_c\mathbf{v}\left(t\right)\text{,}\label{eq:cssx}\\ 
\mathbf{y}\left(t\right)& = & \mathbf{C}_c\mathbf{x}\left(t\right)\text{,} \label{eq:cssy}
\end{IEEEeqnarray}
with state vector~\mbox{$\mathbf{x}=\begin{bmatrix}q_l&i_l\end{bmatrix}^\T$}, where~$q_l$ is the charge of~$i_l$; input~\mbox{$u=i_c$}; measured input disturbance vector~\mbox{$\mathbf{v}=\begin{bmatrix}i_d&v_s\end{bmatrix}^\T$}; and output vector~\mbox{$\mathbf{y}=\begin{bmatrix}v_c&i_l\end{bmatrix}^\T$}, where~$v_c$ is the voltage of the capacitor~$C_2$. The parameters of the model are
\begin{IEEEeqnarray}{rCl}
\IEEEnonumber\IEEEyessubnumber*
\mathbf{A}_c&=&\begin{bmatrix}0 & 1\\-\frac{1}{C_2L_2} &-\frac{R_2+R_1}{L_2}\end{bmatrix}\text{,}\label{eq:Ac}\\
\mathbf{b}_c&=&\begin{bmatrix}0&\frac{R_1}{L_2}\end{bmatrix}^\T\text{,}\label{eq:bc}\\
\mathbf{F}_c&=&\begin{bmatrix}0 & 0\\\frac{R_1}{L_2} & \frac{1}{L_2}\end{bmatrix}\text{,}\label{eq:Fc}\\
\mathbf{C}_c&=&\begin{bmatrix}\frac{1}{C_2} & 0\\0 & 1\end{bmatrix}\text{.}\label{eq:Cc}
\end{IEEEeqnarray}
%%%%%%%%%%%
\subsection{Normal form transformation}\label{ssec:nform}
The kernel cost function described in subsection~\ref{ssec:costkern}, is designed for a second order system where the states are rotating along the circular limit cycle of radius~$\rho$. For simplicity, the radius can be scaled to unity,~\mbox{$\rho=1$}, such that the limit cycle is the unit circle. Therefore, the second order model from subsection~\ref{ssec:circmod} needs to be transformed into normal form, such that its states in undisturbed steady-state conditions reside within the unit circle. This is achieved by the transformation
\begin{IEEEeqnarray}{rCl}
\tilde{\mathbf{x}}& = & \mathbf{M}^{-1}\mathbf{x}\text{,} \label{eq:tx}
\end{IEEEeqnarray}
where
\begin{IEEEeqnarray}{rCl}
\IEEEnonumber\IEEEyessubnumber*
\mathbf{M}&=&\begin{bmatrix}0 & \rho_v C_2\\\rho_i & 0\end{bmatrix}\text{,}\label{eq:tmat}
\end{IEEEeqnarray}
and~\mbox{$\rho_v\!\approx\!1.11$} and~\mbox{$\rho_i\!\approx\!3.49$} are the steady state undisturbed amplitudes of~$v_c$ and~$i_l$ respectively. With this transformation matrix the model parameters are redefined as
\begin{IEEEeqnarray}{rCl}
\IEEEnonumber\IEEEyessubnumber*
\tilde{\mathbf{A}}&=&\mathbf{M}^{-1}\mathbf{A}_c\mathbf{M}\text{,}\label{eq:Au}\\
\begin{bmatrix}\tilde{\mathbf{b}}&\tilde{\mathbf{F}}\end{bmatrix}&=&\mathbf{M}^{-1}\begin{bmatrix}\mathbf{b}_c&\mathbf{F}_c\end{bmatrix}\text{,}\label{eq:buFu}\\
\tilde{\mathbf{C}}&=&\mathbf{C}_c\mathbf{M}\text{.}\label{eq:Cu}
\end{IEEEeqnarray}

This transformed state-space model is then taken to the discrete-time domain using the zero-order-hold method, suitable for a digital controller; leading to the corresponding discrete-time parameters: $\mathbf{A}$, $\mathbf{b}$, $\mathbf{F}$, and~$\mathbf{C}$.
%%%%%%%%%%%
\subsection{Simulation setup}\label{ssec:simset}
The simulation is developed for a power system of frequency~\mbox{$f=\SI{50}{\hertz}$} and supply voltage~\SI{400}{\volt}. The total simulation time is~\SI{0.1}{\second}, considering a sampling time of~$\tau=\SI{0.2}{\milli\second}$ and a prediction horizon of~$H_p=200$.

The limit cycle parameters are:~\mbox{$\mu=\num{e-2}$}, \mbox{$\alpha=-\num{e-2}$}, and~\mbox{$\phi=2\pi f\tau$}.The disturbance current~$i_d$ contains a 3\textsuperscript{rd} and 5\textsuperscript{th} order harmonic distortion with amplitudes of~\SI{2}{\ampere} and~\SI{3}{\ampere}, and phase shifts of~$\arctan\left(\frac{4}{3}\right)$ and~$\arctan\left(\frac{3}{4}\right)+\frac{\pi}{2}$, respectively. For the input Fourier parameterization, the harmonic upper band is~$h=5$.

The controller uses a periodic receding horizon regime to reduce the computation time~\citep{CaPaLi18}. In this strategy the optimal input sequence~$\mathbf{U}^\star$ is calculated once per period instead of per sample; together with the future measured input disturbance prediction~$\mathbf{V}$, which is assumed to be the same as the previous period's measurement. This applies as the compensation is for steady-state periodic distortions. Access to the normal form initial values~$\tilde{\mathbf{x}}_0$, which also defined the starting phase and radius, is assumed.

The solver used was~\texttt{fminunc} from~\textsc{matlab} for general unconstrained minimization. The selected algorithm was~\texttt{quasi-newton}  with default optimality and step tolerances of~\num{e-6}. The average computational time per period was~\SI{0.8}{\second} on an Intel\textsuperscript{\textregistered} Core\texttrademark i7-7700K processor. If constraints would be needed, e.g.\@ physical limitations of the actuator, a different solver setup would be required.
%%%%%%%%%%%
\subsection{Simulation results}\label{ssec:res}
\begin{figure}[ht!]
\centering
\includegraphics[scale=1]{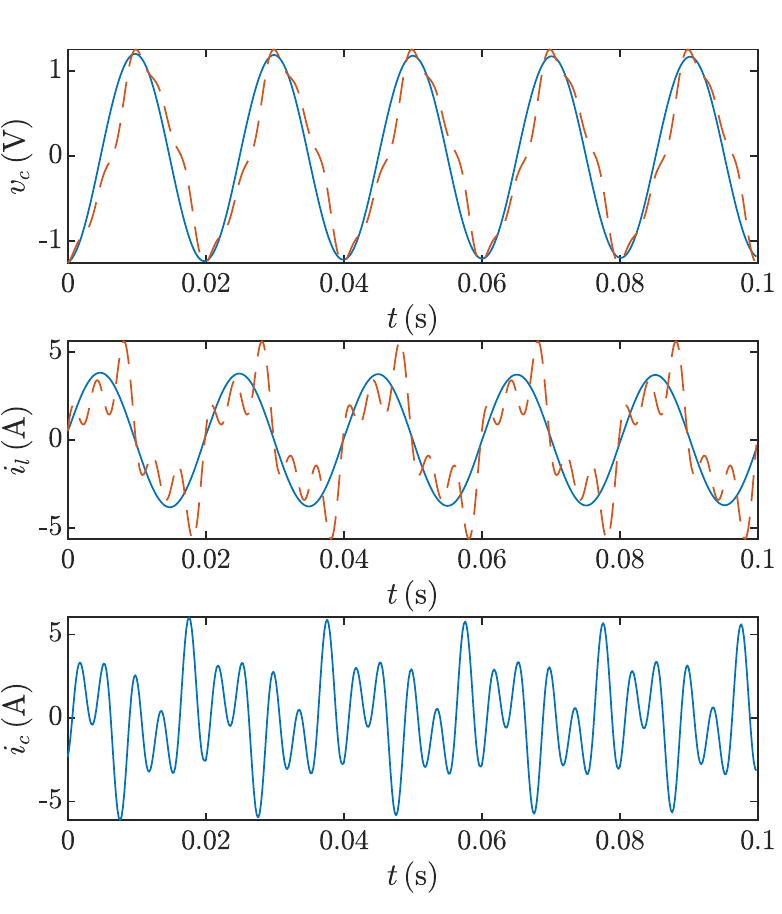}
\caption{Simulation results for the capacitor voltage~$v_c$, load current~$i_l$, and compensation current~$i_c$}
\label{fig:simres}
\end{figure}
In~Fig.~\ref{fig:simres} from top to bottom, the simulation results are shown for the capacitor voltage~$v_c$, the load current~$i_l$, and the~\ac{apf} compensation current~$i_c$. The dashed signals show the reference uncompensated responses, while the solid line signals show the controlled compensation results. Both output signals are successfully compensated to fundamental harmonic shape, lowering the~\ac{thd} from~\SI{15.8}{\percent} to under~\SI{1}{\percent} and from~~\SI{59.8}{\percent} to under~\SI{1}{\percent} for~$v_c$ and~$i_l$ respectively. The~\ac{thd} was calculated against the~\SI{50}{\hertz} fundamental as suggested by~\cite{Sh05}, thus the low~\ac{thd} values confirm that the frequency of the results is close to the fundamental. As quality reference, the results are way below the~\SI{8}{\percent}~\ac{thd} limit defined for voltages in the European Norm~\mbox{EN 50160}~\citep{CE10}.

\begin{figure}[ht!]
\centering
\includegraphics[scale=1]{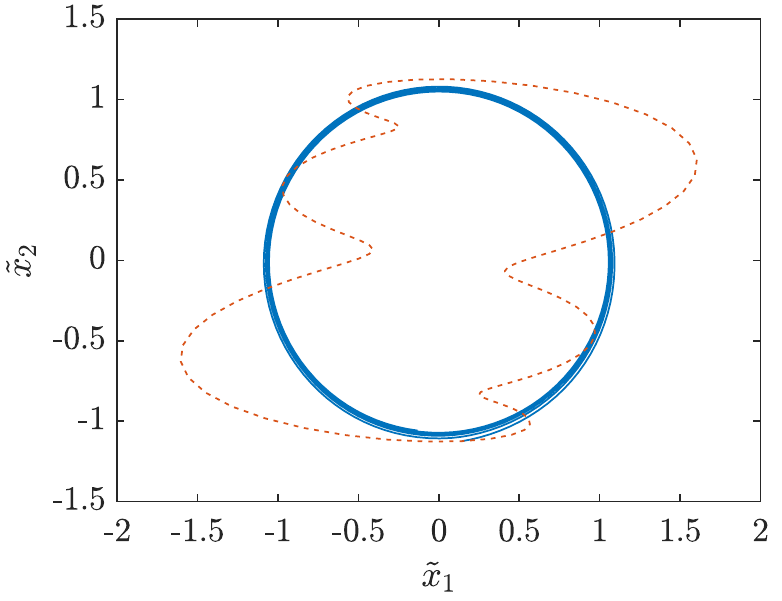}
\caption{Normal form model states phase portrait}
\label{fig:simphaseport}
\end{figure}

In order to analyze the circular limit cycle action of the controller, Fig.~\ref{fig:simphaseport} shows a phase portrait of the simulation results for the normalized model states~\mbox{$\tilde{\mathbf{x}}=\begin{bmatrix}\tilde{x}_1&\tilde{x}_2\end{bmatrix}^\T$}. The dashed line corresponds to the uncompensated system as a reference, while the solid line to the controlled response. Once again, it can be observed how the controller steers the states into the unit circle, effectively imposing the limit cycle dynamics with the right amplitudes.
\section{Conclusion}\label{sec:concl}
The proposed~\ac{lcmpc} approach uses the dynamics of circular limit cycles to define a residual cost for the shape of the state signals. This leads the state signals to a fundamental harmonic shape with a specific frequency and amplitude, given by the parameters of the limit cycle.

In the application example, the controller manages to successfully compensate the~\ac{thd} of the load voltage and current simultaneously to satisfactory levels, while ensuring the correct frequency and amplitude.

Future research focuses on the cost function convexity, and analytic Hessian and gradient formulations for lower computational time. Comparisons with other controllers, as well as enhanced future disturbance and phase estimation methods, are to be explored.

This work was developed as part of the Ph.D.\@ research project from Carlos Cateriano Yáñez under the direct supervision of Gerwald Lichtenberg, and in close collaboration with Georg Pangalos and Javier Sanchis Sáez.

%===============================================================================

%\input{tex/ack}
\bibliography{bib/LCMPC}

\begin{thebibliography}{13}
\providecommand{\natexlab}[1]{#1}
\providecommand{\url}[1]{\texttt{#1}}
\providecommand{\urlprefix}{URL }
\expandafter\ifx\csname urlstyle\endcsname\relax
  \providecommand{\doi}[1]{doi:\discretionary{}{}{}#1}\else
  \providecommand{\doi}{doi:\discretionary{}{}{}\begingroup
  \urlstyle{rm}\Url}\fi

\bibitem[{Brandonisio and Kennedy(2014)}]{BrKe14}
Brandonisio, F. and Kennedy, M.P. (2014).
\newblock \emph{Noise-{{Shaping All}}-{{Digital Phase}}-{{Locked Loops}}:
  {{Modeling}}, {{Simulation}}, {{Analysis}} and {{Design}}}.
\newblock Analog {{Circuits}} and {{Signal Processing}}. {Springer
  International Publishing}.
\newblock \doi{10.1007/978-3-319-03659-5}.

\bibitem[{Cateriano~Y{\'a}{\~n}ez et~al.(2018)Cateriano~Y{\'a}{\~n}ez,
  Pangalos, and Lichtenberg}]{CaPaLi18}
Cateriano~Y{\'a}{\~n}ez, C., Pangalos, G., and Lichtenberg, G. (2018).
\newblock An {{Approach}} to linear state {{Signal Shaping}} by quadratic
  {{Model Predictive Control}}.
\newblock In \emph{2018 {{European Control Conference}} ({{ECC}})}, 1--6.
  {IEEE}, {Limassol, Cyprus}.
\newblock \doi{10.23919/ECC.2018.8550379}.

\bibitem[{CENELEC(2010)}]{CE10}
CENELEC (2010).
\newblock {{EN}} 50160: {{Voltage}} characteristics of electricity supplied by
  public distribution systems.
\newblock Technical report, {European Committee for Electrotechnical
  Standardization}.

\bibitem[{Ding and Zhang(2009)}]{DiZh09}
Ding, Y. and Zhang, Q. (2009).
\newblock Simplest normal forms of generalized {{Neimark}}-{{Sacker}}
  bifurcation.
\newblock \emph{Transactions of Tianjin University}, 15(4), 260--265.
\newblock \doi{10.1007/s12209-009-0046-x}.

\bibitem[{Egerstedt and Hu(2001)}]{EgXi01}
Egerstedt, M. and Hu, X. (2001).
\newblock Formation constrained multi-agent control.
\newblock \emph{IEEE Transactions on Robotics and Automation}, 17(6), 947--951.
\newblock \doi{10.1109/70.976029}.

\bibitem[{Guckenheimer and Holmes(1983)}]{GuHo83}
Guckenheimer, J. and Holmes, P. (1983).
\newblock \emph{Nonlinear {{Oscillations}}, {{Dynamical Systems}}, and
  {{Bifurcations}} of {{Vector Fields}}}.
\newblock {Springer New York}, {New York, NY}.
\newblock \doi{10.1007/978-1-4612-1140-2_1}.

\bibitem[{Ivancevic and Ivancevic(2007)}]{IvIv07}
Ivancevic, V.G. and Ivancevic, T.T. (2007).
\newblock \emph{Computational Mind: A Complex Dynamics Perspective}, volume~60
  of \emph{Studies in {{Computational Intelligence}}}.
\newblock {Springer-Verlag Berlin Heidelberg}, first edition.
\newblock \doi{10.1007/978-3-540-71561-0}.

\bibitem[{Kuznetsov(1998)}]{Ku98}
Kuznetsov, Y.A. (1998).
\newblock \emph{Elements of {{Applied Bifurcation Theory}}}, volume 112 of
  \emph{Applied {{Mathematical Sciences}}}.
\newblock {Springer-Verlag New York}, {New York, NY}, second edition.

\bibitem[{Liang(2017)}]{Li17}
Liang, X. (2017).
\newblock Emerging {{Power Quality Challenges Due}} to {{Integration}} of
  {{Renewable Energy Sources}}.
\newblock \emph{IEEE Transactions on Industry Applications}, 53(2), 855--866.
\newblock \doi{10.1109/TIA.2016.2626253}.

\bibitem[{Maciejowski(2002)}]{Ma02}
Maciejowski, J.M. (2002).
\newblock \emph{Predictive {{Control}} with {{Constraints}}}.
\newblock {Pearson education}.

\bibitem[{Rao et~al.(2008)Rao, Mishra, and Ghosh}]{RaMiGh08}
Rao, U.K., Mishra, M.K., and Ghosh, A. (2008).
\newblock Control {{Strategies}} for {{Load Compensation Using Instantaneous
  Symmetrical Component Theory Under Different Supply Voltages}}.
\newblock \emph{IEEE Transactions on Power Delivery}, 23(4), 2310--2317.
\newblock \doi{10.1109/TPWRD.2008.923053}.

\bibitem[{Shmilovitz(2005)}]{Sh05}
Shmilovitz, D. (2005).
\newblock On the definition of total harmonic distortion and its effect on
  measurement interpretation.
\newblock \emph{IEEE Transactions on Power Delivery}, 20(1), 526--528.
\newblock \doi{10.1109/TPWRD.2004.839744}.

\bibitem[{Weihe et~al.(2020)Weihe, Cateriano~Y{\'a}{\~n}ez, Pangalos, and
  Lichtenberg}]{WeCaPa20}
Weihe, K., Cateriano~Y{\'a}{\~n}ez, C., Pangalos, G., and Lichtenberg, G.
  (2020).
\newblock Constrained {{Linear State Signal Shaping Model Predictive Control}}
  for {{Harmonic Compensation}} in {{Power Systems}}.
\newblock In \emph{21st {{IFAC World Congress}}}. {Berlin, Germany}.

\end{thebibliography}

\end{document}